\documentclass{aa}
\usepackage{graphicx}

\begin{document}

\voffset -0.5in

\title{Eclipsing binary millisecond pulsar PSR J1740-5340 -- evolutionary considerations and
observational test}

\author{Ene Ergma\inst{1} \and Marek J. Sarna\inst{2}}

\offprints{E. Ergma}

\institute{
           Physics Department, Tartu University, {\"U}likooli 18, 50090
           Tartu, Estonia\\
           \email{ene@physic.ut.ee}
\and
           N. Copernicus Astronomical Center, Polish Academy of Sciences,
           ul. Bartycka 18, 00--716 Warsaw, Poland.\\
           \email{sarna@camk.edu.pl}
}

\date{received March 20;   accepted}

\abstract {We perform evolutionary calculations for a binary
system with initial parameters: $M_{sg,i}$= 1$M\odot$ and
$M_{ns,i}$= 1.4 $M\odot$ and $P_{orb,i}(RLOF)$= 1.27 d to produce
observed binary parameters for the PSR J1740--5340. Our
calculations support model proposed by D'Amico et al. (2001) in
which this binary may be progenitor of a millisecond pulsar +
helium white dwarf system. We propose observational test to verify
this hypothesis. If the optical companion lack of carbon lines in
its spectrum shows but the oxygen and nitrogen lines are present
then our model correctly describes the evolutionary stage of PSR
J1740--5340.  }

\authorrunning{Ene Ergma and M.J. Sarna}
\titlerunning{Eclipsing BMSP PSR J1740--5340...}
\maketitle

\section{Introduction}

The millisecond pulsar PSR J1740--5340 was discovered during a
systematic search of the Galactic globular cluster using the
Parkers radio telescope (D'Amico et al. 2001 a, b). The PSR
J1740--5340 in the globular cluster NGC 6397 is a member of a
binary system with a relatively wide orbit of period 1.35 days, a
companion with mass $M_c > 0.19 ~M_\odot$, and is eclipsed for
about 40\% of its orbit at a frequency of $\nu$=1.4 GHz.
Spin--down age of PSR J1740--5340 ($\tau_{sd}$=$P_{pul}/2{\dot
P}_{pul}$) is $\sim $350 Myr and its surface magnetic field is B =
3.2$~ 10^{19} (P_{pul} {\dot P}_{pul})^{1/2}$ $\sim 8 ~ 10^8~$G.
Ferraro et al. (2001) report the optical identification of the
companion to the millisecond pulsar. They found that observed
modulation is just $\sim$  0.2 mag; that can be reproduced only if
the companion has almost filled its Roche lobe and the orbital
plane is nearly edge--on ({\it i} $\sim 90^0$). These two
requirements are accomplished by a companion of mass in the range
0.19--0.22$~M_\odot$, whose Roche lobe radius
(1.32--1.42$~R_\odot$) just matches the lower limit for observed
radius $R_c \sim$ 1.3--1.8$~R_\odot$. D'Amico et al.(2001 a, b)
and Ferraro et al. (2001) propose two hypothesis about the origin
of this system.

For explaining the large mass loss rate they proposed that the
companion is an evolved star that spun up the pulsar to
millisecond period. The mass accretion and spinning up of the
pulsar would now be inhibited by the pulsar wind flux which could
expel the matter overflowing from the Roche lobe of the companion.
This would be the first confirmed example of a recently born
millisecond pulsar.

Another hypothesis suggested that the optical companion is a
Main--Sequence (M--S) star perturbed by the energetic flux emitted
from the millisecond pulsar. The only 5\% of the impinging power
would be needed to sustain the inferred mass loss rate ${\dot M} <
2~10^{-11} ~M_\odot ~yr^{-1}$ (D'Amico et al. 2001 b). This model
should be more easily applicable to a M--S star, as large
convective envelope favours bloading of the star.

Ergma, Sarna \& Antipova (1998) investigated evolution and
formation of short period binary millisecond pulsars near the so
called bifurcation period (Pylyser \& Savonije 1988). They found
that if Roche lobe filling of the secondary  occurs near
bifurcation period  then it is possible that the final orbital
period (after detachment from Roche lobe) is near 1--2 days.

In this research note we shall show that it is possible to produce
observational parameters for the system PSR J1740--5340 and that
hypothesis proposed by D'Amico et al. (2001 a) really may work. If
so we are observing binary right  before formation of the detached
He white dwarf + millisecond pulsar system.  We also propose
observational test to distinguish between the two hypothesis (M--S
star vs. evolved star).

\section{The evolutionary code}

The evolutionary sequences we have calculated are comprised of
two main phases:

\noindent
(i) detached evolution lasting until the companion
fills its Roche lobe,

\noindent (ii) semi--detached evolution.

\subsection{Detached phase}

The duration
of the detached phase is somewhat uncertain;
it may be determined either by the nuclear time--scale or by the
much shorter time--scale of the orbital angular momentum loss owing
to the magnetized stellar wind.

\subsection{Semi--detached phase}

In our calculations we assume that the semi--detached evolution of
a binary system is non--conservative, i.e. the total mass and
angular momentum of the system are not conserved. The formalism
which we have adopted is described in Muslimov \& Sarna (1993,
1995). We introduce the parameter, $f_1$, characterizing the loss
of mass from the binary system and defined by the relations

\begin{equation}
\dot M = \dot M_{sg} f_1 ~~~~and~~~~ \dot M_{ns}
= - \dot M_{sg} (1 - f_1) ,
\end{equation}

where $\dot M$ is the mass--loss rate from the system, $\dot
M_{sg}$ is the rate of mass--loss from the donor (secondary) star,
and $\dot M_{ns}$ is the accretion rate onto the neutron star
(primary). The matter leaving the system will carry away its
intrinsic angular momentum in agreement with formula

\begin{equation}
{{\dot J} \over {J}} = f_2 {{M_{ns} \dot M} \over {M_{sg} M}}
~~~~~yr^{-1}  ,
\end{equation}

where M, $M_{ns}$ and $M_{sg}$ are the total mass of the system,
the masses of the neutron star and donor star respectively. Here
we have introduced the additional parameter, $f_2$, which
describes the efficiency of the orbital angular momentum loss from
the system due to a stellar wind (Tout \& Hall 1991). In our
calculations we assume $f_1$=1 (all transferred matter loss from
the system) and $f_2$=1 (typical stellar wind -- see Tout \& Hall
1991).

We also assume that the donor star, possessing a convective
envelope, experiences magnetic braking (Mestel 1968, Mestel \&
Spruit 1987), and as a consequence of this, the system loses its
orbital angular momentum.

In the semi--detached phase we have also included the effect of illumination
of the donor
star by the millisecond pulsar. In our calculations we assume that the
illumination of the component by the hard (X--ray and $\gamma$--ray)
radiation from the pulsar leads to additional heating of its
photosphere (Muslimov \& Sarna 1993). The effective temperature,
$T_{eff} $, of the companion during the illumination stage is
determined from the relation

\begin{equation}
L_{in} + L_{ill} = 4 \pi \sigma R_{sg}^2 T^4_{eff} ,
\end{equation}

where $L_{in} $ is the intrinsic luminosity corresponding to the
radiation flux coming from the stellar interior $\sigma $ is the
Stefan--Boltzmann constant and $ R_{sg} $ is the stellar radius.
$L_{ill} $ is the millisecond pulsar radiation that heats the
photosphere, and is

\begin{equation}
L_{ill}= f_3 \left(\frac{R_{sg}}{2a}\right)^2 L_{rot}
\end{equation}

and $L_{rot}$ is ``rotational luminosity'' of the neutron star due to
magneto--dipole radiation (plus a wind of relativistic particles)

\begin{equation}
L_{rot}=\frac{2}{3c^3} B^2 R_{ns}^6\left(\frac{2\pi}{P_{pul}}\right)^4
\end{equation}

where $R_{ns}$ is the neutron star radius, B is the value of the
magnetic field strength at the neutron star and $P_{pul}$ is the
pulsar spin period. $f_3$ is a factor characterizing the
efficiency of transformation of irradiation flux into thermal
energy (in our case we take $f_3 = 2~ 10^{-3}$). Note that in our
calculations the effect of irradiation is formally treated by
means of modification of the outer boundary condition, according
to above the relation.

\subsection{Pulsar spin evolution}

We also follow the rotational evolution of the neutron star which
is determined mainly by the accretion rate and the temporal
evolution of the magnetic field of the neutron star. For the
temporal evolution of the magnetic field the model by Urpin \&
Muslimov (1992) has been used.

As the neutron star spins up, the net accretion torque gradually
decreases and the spin period reaches its equalibrium value given
(see e.g. van den Heuvel 1977) by

\begin{equation}
P_{eq}= 2.1 B_9^{6/7}\left({{{\dot M}_{sg}} \over {{\dot
M}_{edd}}} \right)^{-3/7} ms
\end{equation}

When the spin period of the neutron star reaches $P_*$, determined
as

\begin{equation}
P_*= 1.9 ~ B^{1/2}_9 \left({{{\dot M}_{sg}} \over {{\dot
M}_{edd}}} \right)^{-1/4}
\left({{R_{sg}}\over{M_{sg}}}\right)^{1/8} ms
\end{equation}

where $B_9$= B/$10^9$G, ${\dot M}_{edd}$ is the Eddington mass
accretion rate (for more details see Muslimov \& Sarna, 1993),
then accreted matter is expected to be ejected from the system by
the pressure of magneto--dipole radiation.

\subsection{Nuclear network}

Our nuclear reaction network is based on that of Kudryashov \&
Ergma (1980), who included the reactions of the CNO tri--cycle in
their calculations of hydrogen and helium burning in the envelope
of an accreting neutron star. We have included the reactions of
the proton--proton (PP) chain. Hence we are able to follow the
evolution of the nuclei: $^{1}$H, $^{3}$He, $^{4}$He, $^{7}$Be,
$^{12}$C, $^{13}$C, $^{13}$N, $^{14}$N, $^{15}$N, $^{14}$O,
$^{15}$O, $^{16}$O, $^{17}$O and $^{17}$F. We assume that the
abundances of $^{18}$O and $^{20}$Ne stay constant throughout the
evolution. We use the reaction rates from: Fowler, Caughlan \&
Zimmerman (1967, 1975), Harris at al. (1983) and Pols et al.
(1995).

\section{Results of calculations}

We  perform calculations for the following system parameters:
$M_{sg,i}$=1$M_\odot$, $M_{ns,i}$=1.4 $M_\odot$,
$P_{orb,i}(RLOF)$= 1.1--1.5 d and chemical composition: X=0.7,
Z=0.003. In Fig. 1 we show the mass--loss rate from the donor star
versus secondary mass. If Roche lobe filling of secondary occurs
near bifurcation period (the bifurcation period separates the
formation of the converging systems: $P_f$$<$$P_{orb,i}(RLOF)$
from the diverging systems: $P_f$$>$$P_{orb,i}(RLOF)$, where $P_f$
and $P_{orb,i}(RLOF)$ are the final and initial orbital periods,
respectively) then as it was already found by Tutukov et al.
(1985) the orbital evolution is quite  different then in the case
when the secondary Roche lobe filling occurs far from bifurcation
period. At the beginning, mass loss rate is high and the pulsar
will spin up  to current spin period, then the mass loss rate
quickly decreases. Near bifurcation period two time scales:
nuclear time--scale of the secondary and the angular momentum loss
time--scale are very close each others. Therefore, during the
binary evolution orbital period changes are insignificant in
comparison with initial orbital period (Tutukov et al. 1985, Ergma
et al. 1998).

\begin{figure}
\centering
\includegraphics[width=8cm]{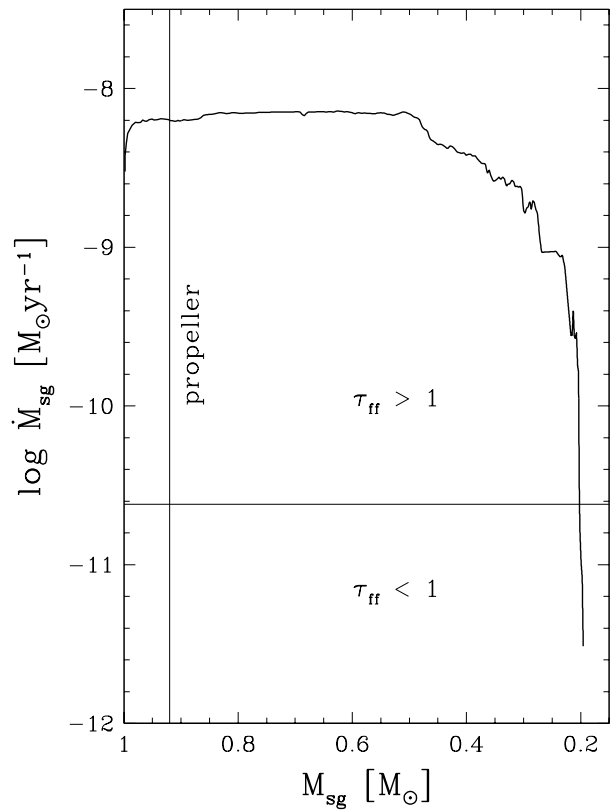}
\caption{The evolution of the mass transfer rate versus secondary
 mass. The horizontal line below which $\tau_{ff}$  is less than 1 is
shown. To the right from vertical line the pulsar is in
``propeller phase''}
\end{figure}

For various accretion rate we can estimate the optical depth of
the stellar wind due to free--free absorption (Illarionov \&
Sunyaev 1975)

\begin{equation}
\tau_{ff}={100 ({{{\dot M}_{sg}}/{{\dot
M}_{edd}}})^2\over{(M_{sg}+M_{ns})/M_\odot}}
\left(\frac{\lambda}{75 cm}\right)^2
\left(\frac{T}{10^4}\right)^{-1.5} \left(\frac{P}{1yr}\right)^{-2}
\end{equation}

In Fig 1. for secular mass loss rate the line below  $\tau_{ff}$
is less than unity and the region where pulsar is in ``propeller
phase'' (right from vertical line) are shown. Although the optical
depth for secular mass loss is less than unity due to ``propeller
phase'' plasma inside the system may be very clumsy with much
higher optical depth and free--free absorption alone could be
viable eclipse mechanism.

\begin{figure}
\centering
\includegraphics[width=8cm]{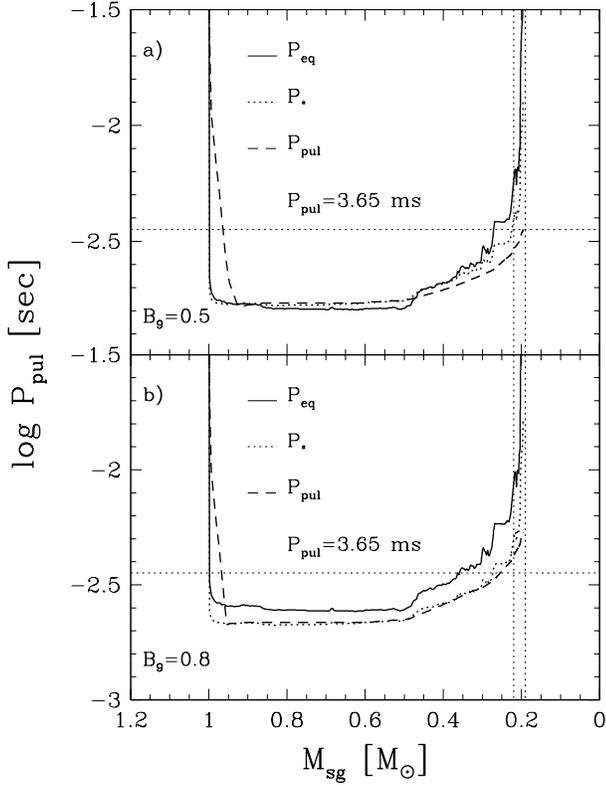}
\caption{The evolution of the spin period ($P_{pul}$) of the
pulsar, equilibrium period ($P_{eq}$) and  $P_*$ versus the mass
of the secondary for two values of magnetic field strength 5$~
~10^8~$ G (a) and 8$~10^8~$G . The horizontal dotted line shows
the spin period of PSR J1740--53. The vertical lines show the mass
range of the secondary star.}
\end{figure}

\begin{figure}
\centering
\includegraphics[width=8cm]{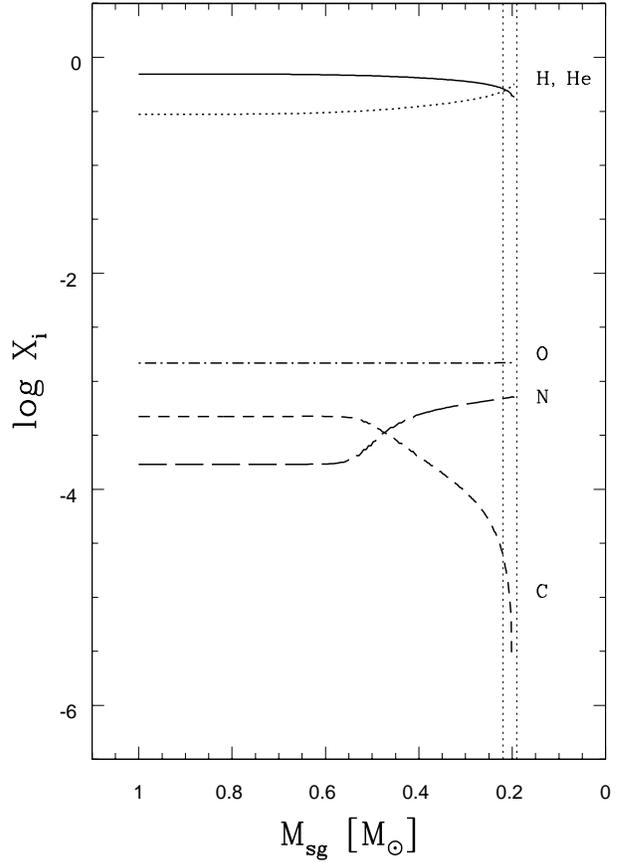}
\caption{The evolution of the red giant surface abundances  of H,
He, C, N and O as a function of the subgiant mass $M_{sg} $. The
vertical lines show the mass range of the secondary star.}
\end{figure}

The model which provides the best fitting for the observed
physical and orbital parameters is for $P_{orb,i}$=1.27 d. During
semi--detached evolution, near $P_{orb}$ =1.35 days secondary mass
has decreased to $M_{sg}$ $\sim$ 0.2 $M\odot$, its radius is
$\sim$ 1.38$R\odot$ and effective temperature 5620 K. Both values
fit well with observed values (D'Amico et al. 2001 a, Ferraro et
al. 2001). From Taylor et al. (2001) we know that $m_v$=16.66 mag
for optical component of PSR J1740+5340. From Harris (1996) we
find that distance modulus to NGC 6397 is 12.25 mag. These numbers
gives absolut maginitude $M_v$=4.41 mag. From our theoretical
temperature and radius determination we calculated $M_{bol}$=
4.166 mag. The bolometric correction from Houdashelt, Bell \&
Sweigert (2000) gives BC=0.272$\pm$0.007 for assuming metalicity
between (log[Fe/H]) --2 and --3. The theoretical value of
$M_v$=4.44 mag agree perfectly with Taylor at al. (2001)
determination.

Fig. 2 a) and b) depict the temporal behaviour of the neutron star
spin period, $P_{pul} $, as well as $P_{eq} $ and $P_{*} $, in the
case when B=$5~10^8 $ and $8~10^8 $ G, respectively, at the
beginning of the accretion stage. Fig. 2 shows that, prior to
reaching its equilibrium value before the spin period reaches its
equilibrium value ($P_{eq} $), the spin period decreases to the
value of $P_* $ and the condition $P_{pul} \leq P_* $ is
satisfied. After this, the evolution of the neutron star spin
period is determined by magneto--dipole braking. When the spin
period of the neutron star reaches $P_*$ (eq. 6) then the accreted
matter is expected to be ejected from the system by the pressure
of magneto--dipole radiation. It happens when $M_{sg} \sim 0.92
~M_\odot$, ${\dot M}_{sg} \sim 6~ 10^{-9} ~M_\odot ~yr^{-1}$ for
B=$5~10^8 ~$G and when $M_{sg} \sim 0.95 ~M_\odot$, ${\dot M}_{sg}
\sim 6.4~ 10^{-9} ~M_\odot ~yr^{-1}$ for B=$8~10^8 ~$G,
respectively. It should be pointed out that, in both cases the
mass of the neutron star will grow, with ejection of accreted
matter stopping at the value $< 1.5 ~M_\odot $ at the beginning of
the ejection phase.

Note also, that for binary system parameters as above, pulsar spin
period is 3.5 and 5 ms for 5$~10^8$ and 8$~10^8 ~$G, respectively.
This will suggest that the value of 8$~10^8~$G given by D'Amico et
al. (2001 b) is overestimated.

\section{Observational test}

We propose that observations of abundances of  C, N, O elements in
the spectra of optical companion of PSR J1740--5340 may give
additional information about  the nature of the secondary being
either i) perturbed low--mass M--S star or ii) evolved star
(helium) which has lost its envelope during the accretion phase.
In Fig. 3 we present evolution of the optical companion surface
chemical composition as a function of the secondary mass (for case
ii)). Near orbital period $P_{orb}$= 1.35 d for the evolved star,
carbon and oxygen are depleted by --2.6 dex and --0.8 dex (in
comparison to cosmic abundances), respectively. Only nitrogen
lines will show cosmic abundances (see Fig. 3). In the case of the
low--mass M--S star C, N, O abundances will be as for the halo and
disc metal poor ([Fe/H] $<$ -1.5) dwarfs (Carretta, Graton \&
Sneden 2000 and references therein). In the case of the evolved
star (helium) N will be overabundant (approxymately solar) when O
abundance will be as for metal poor dwarfs. It will be lack of
carbon lines in the evolved star spectrum.

\section{Conclusions}

We accept  D'Amico et al. (2001 a) hypothesis that the eclipsing
binary millisecond pulsar PSR J 1740--5340 may be regarded as a
recently born binary millisecond pulsar system. Our calculations
shows that it is possible to build  the evolutionary scenario for
the binary system with the orbital parameters, secondary mass and
radius which fit well those for PSR J1740--5340. We also suggest
that the observed abundances of C, N, O elements in the spectra of
optical companion of millisecond pulsar can tell us about
evolutionary stage of secondary. If carbon is depleted then we
really have a unique situation involving pre--He wd + millisecond
pulsar system.

\section*{\sc Acknowledgments}

This work is partly supported through grants 2--P03D--005--16 of the
Polish National Committee for Scientific Research  and Collaborative
Linkage Grant PST.CLG.977383.  EE acknowledge support through Estonian SF grant 4338.

\end{document}